\documentclass[twocolumn,amsmath,amssymb,floatfix,superscriptaddress]{revtex4-1}
\usepackage{graphicx}
\usepackage{amssymb}
\usepackage{amsmath}
\usepackage{color}
\usepackage{ wasysym }
\usepackage{hyperref}

\def\barray{\begin{array}}
\def\earray{\end{array}}
\def\be{\begin{equation}}
\def\ee{\end{equation}}
\def\ben{\begin{equation} \nonumber}
\def\een{\end{equation}}
\def\ban{\begin{eqnarray*}}
\def\ean{\end{eqnarray*}}
\def\ba{\begin{eqnarray}}
\def\ea{\end{eqnarray}}

\def\({\left(}
\def\){\right)}

\graphicspath{{./fig/}}

\begin{document}
\title{Tachyon warm inflation with  the effects of Loop Quantum Cosmology in the light of Planck 2015
}

\author{Vahid  Kamali}
\email{vkamali@basu.ac.ir}
\affiliation{Department of Physics, Bu-Ali Sina University, Hamedan 65178,
016016, Iran}
\author{Spyros Basilakos}
\email{svasil@academyofathens.gr}
\affiliation{ Academy of Athens, Research Center for Astronomy \& Applied 
Mathematics, Soranou Efessiou 4, 11-527, Athens, Greece}
\author{ Ahmad  Mehrabi}
\email{Mehrabi@basu.ac.ir}
\affiliation{Department of Physics, Bu-Ali Sina University, Hamedan 65178,
016016, Iran}
%\thankstext[$\star$]{t1}{Thanks to the title}
%\thankstext{e}{e-mail: vkamali@basu.ac.ir}
%\thankstext{e1}{e-mail:svasil@academyofathens.gr}
%\thankstext{e2}{e-mail: Mehrabi@basu.ac.ir}
%\institute{Department of Physics, Bu-Ali Sina University, Hamedan
%65178, 016016, Iran\label{addr}
%         \and
  %        Academy of Athens, Research Center for Astronomy \& Applied 
  %Mathematics, Soranou Efessiou 4, 11-527, Athens, Greece\label{addr1}
          %\and
          %\emph{Present Address:} Street, City, Country\label{addr3}
%}
\author{Meysam Motaharfar}
\email{mmotaharfar2000@gmail.com}
\affiliation{Department of Physics, Shahid Beheshti University, G. C., Evin,Tehran 19839, Iran}
\author{Erfan Massaeli}
\email{erfan.massaeli@gmail.com}
\affiliation{Department of Physics, Shahid Beheshti University, G. C., Evin,Tehran 19839, Iran}
\date{\today}

%\maketitle
%========================================================================
%========================================================================
%{\bf Abstract:}
\begin{abstract}
We investigate the observational signatures of quantum cosmology 
in the Cosmic Microwave Background data provided by Planck collaboration.
We apply the warm inflationary paradigm with a tachyon scalar field 
to the loop quantum cosmology. In this context, we first provide  
the basic cosmological functions in terms of the tachyon field. 
We then obtain the slow-roll parameters and the 
power spectrum of scalar and tensor fluctuations respectively.
Finally, we study the performance of various warm inflationary scenarios
against the latest Planck data and we find a family 
of models which are in agreement with the observations. 
\end{abstract}
%\keywords{Warm Inflation, Tachyon Field, Loop Quantum Cosmology (LQC), CMB Data,  Slow-Roll Approximation. }
%\ccode{PACS numbers: 98.80.Cq}
\maketitle
\section{Introduction}
%\section{Introduction}
Recent studies of %photons in 
the Cosmic Microwave Background (hereafter 
CMB) %radiation
have placed tight constraints on
single scalar-field models of slow-roll inflation. Specifically, 
based on Planck data \cite{Ade:2015lrj} it has been 
found that the inflationary models which are in agreement with the data 
are those with very low tensor-to-scalar fluctuation ratio $
r = P_T/P_s \ll 1$, with a scalar spectral index $n_s \simeq 0.96 $. Actually, the upper bound found by Planck 
team~\cite{Ade:2015lrj}, on this ratio, as a result of the
non-observation of B-modes, is $r < 0.11$.

% which implies that 
% $H\lesssim 10^{-5}M_p$, where 
%$M_{p}=\frac{1}{\sqrt{8\pi G}} \simeq 2.43 \times 10^{18}$Gev 
%$M_{p} \simeq 1.22 \times 10^{19}$Gev 
%is the reduced 
%Planck mass.

In the standard picture of 
slow-roll inflation \cite{Guth:1980zm, Albrecht:1982wi}
the kinetic energy (${\dot \phi}^{2}/2$) and the potential energy $V(\phi)$
of the inflaton field satisfies 
${\dot \phi}^{2}/2 \ll V(\phi)$. The latter condition imposes the
de Sitter expansion in the early universe, 
since the corresponding equation of state parameter 
$w\equiv \frac{(\dot{\phi}^{2}/2)-V(\phi)}{(\dot{\phi}^{2}/2)+V(\phi)}$ 
tends to -1. Moreover, in the standard inflation 
possible interactions among other fields with the inflaton must be neglected. 
Subsequently, after the slow-roll era the potential energy 
becomes comparable to the 
kinetic energy. This simply means that we are dealing with the 
so called reheating period in which 
the inflaton starts to oscillate around the 
minimum of the potential and progressively the universe is filled by 
radiation \cite{Shtanov:1994ce,Kofman:1997yn}. 

In the literature one may find other theoretical approaches in order 
to treat the nature of the early universe. Specifically, over the 
last two decades a lot of attention has been paid on the so-called warm 
inflationary pattern. Unlike to standard inflation, 
in this scenario the inflaton field is allowed to interact 
with other light fields, implying that  
radiation production occurs during the slow-roll period and hence
reheating is avoided \cite{Berera:1995ie,Berera:1996fm}. 
Obviously, the main idea of the warm inflationary model 
is quite different with that of the 
standard cold inflation. Indeed,  
warm inflation satisfies 
the condition $T>H$, where $T$ is the temperature 
and $H$ is the Hubble parameter, which implies that
the fluctuations of the inflaton field are thermal instead of fluctuations of the ground state.
These thermal fluctuations play an eminent 
role in large scale structure studies because they are the 
initial seeds of density perturbations
\cite{Hall:2003zp,Moss:1985wn,Berera:1999ws}. 

From the scalar-field viewpoint, warm inflation is characterized 
by a tachyon scalar field with positive potential $V(\phi)$, while 
the corresponding kinetic energy does not obey the standard 
form (k-inflation \cite{ArmendarizPicon:1999rj}).
Also the potential energy has two special properties, 
namely a maximum at $\phi\rightarrow 0$ and a minimum when 
$\phi\rightarrow \infty$. 
For more details regarding 
the warm tachyon inflationary model we refer the reader to 
Refs.\cite{Herrera:2006ck,Setare:2012fg,Deshamukhya:2009wc}. 
In this framework,  
it is well known that under specific conditions 
tachyon fields which are related with
unstable D-branes \cite{Sen:2002nu,
%Das:2008gi,Choudhury:2015hvr,Das:2008af,
Chingangbam:2004ng} provide cosmic acceleration  
\cite{Sen:2002an, Sami:2002fs, ArmendarizPicon:1999rj,Mazumdar:2001mm,Deshamukhya:2009wc,Panda:2005sg,Chingangbam:2004ng,Choudhury:2002xu}
during early times.

In the current article we present 
the dynamical behavior of the effective Loop Quantum Gravity (LQG) theory 
via the Hubble expansion,
and investigate the compatibility of this scenario
with warm tachyon-like inflationary scenarios.
The basic techniques of LQG \cite{Thiemann:2002nj,Bojowald:2006da,Ashtekar:2004eh,Singh:2005xg}, 
%which are resulting
%non-perturbative background independent approach to quantizing
%gravity \cite{Thiemann:2002nj,Bojowald:2006da,Ashtekar:2004eh,Singh:2005xg}, 
can be applied in homogeneous and isotropic space-times 
in order to build a 
%which is known as 
Loop Quantum Cosmology (LQC).
The layout of the paper is as follows: 
At the beginning of Sec. II we present the main points of the
LQC and then we discuss the scalar field description of the 
warm tachyon inflationary paradigm
in the context of LQC. 
In Sec. III we estimate the slow roll parameters and 
in Sec, IV we calculate the temperature at the end of warm inflation.
In Sec V we test the performance of the slow-roll predictions 
against the latest \textit{Planck 2015} data and 
finally, we summarize our conclusions in Sec. VI.

\section{ Loop Quantum Cosmology versus Warm Inflation:}
%\section{Loop Quantum Cosmology versus Warm Inflation}
In this section we briefly present the basic cosmological features
of Loop Quantum Cosmology (LQC). 
In the literature canonical quantization of gravity is given in 
terms of the so called Ashtekar-Barbero
connection variables (see \cite{Ashtekar:2003hd}). 
%{\bf PUT THIS REF IN THE LIST SEE AT THE END OF THE PAPER AND OTHER REFS}). 
Without wanting to enter into the 
full details the phase space of
classical general relativity can be spanned by conjugate
variables $A_q^i$ (connection) and $E_i^q$ (triad) on a
$3-manifold$ $\mathcal{M}$ which encodes curvature and spatial
geometry respectively (labels $q$ and $i$ denote internal indices
of $SU(2)$ and space index respectively). At the cosmological level (LQC)
due to the isotropic and homogeneous symmetries the phase space is
characterized by a single
connection $c$ and a single triad $p$.
 Notice, that the
Poisson bracket of LQC variables is given by 
\begin{equation}\label{}
\nonumber \{c,p\}=\frac{8\pi\gamma}{3M_p^2}
\end{equation}
where $\gamma \simeq 0.2375$ is the dimensionless Barbero-Immirzi parameter, 
derived from the black hole thermodynamics \cite{Ashtekar:2000eq, Corichi:2009wn, Agullo:2010zz, Meissner:2004ju}.
Considering a spatially flat Friedmann-Robertson-
Walker (FRW) metric,
the LQC variables become 
\begin{equation}\label{}
\nonumber c=\gamma\dot{a}~~~~~~~p=a^2,
\end{equation}
in the classical regime,
where $a(t)$ is the scale factor of the universe
and the overdot denotes derivative with respect to 
cosmic time $t$. In the variable system $\{c,p\}$
Ashtekar et al. \cite{Ashtekar:2003hd}
proposed that the 
classical Hamiltonian constraint is given by
\begin{equation}\label{}
\nonumber
\mathcal{H}_{cl}=-\frac{3\sqrt{p}}{\gamma^2}+c^2+\mathcal{H}_m
\end{equation}
where $\mathcal{H}_m$ is the matter Hamiltonian. 
However, in the semi-classical regime  
various authors \cite{Singh:2005xg,Ashtekar:2006wn,Ashtekar:2006uz}
introduced an effective theory of loop
quantum gravity 
which is appropriate for cosmology. In this case the effective 
Hamiltonian constraint is written as 
\begin{equation}\label{}
\nonumber
\mathcal{H}_{\rm eff}=-\frac{3\sqrt{p}}
{\gamma\overline{\mu}^2}\sin^2(\mathcal{\overline{\mu}}c)+\mathcal{H}_m\\
\end{equation}
%\begin{equation}\label{}
%\nonumber
%\mathcal{H}=-\frac{3\sqrt{\textbf{p}}}{\gamma^2}+\textbf{c}^2+\mathcal{H}_{m}
%\end{equation}
where the constant $\overline{\mu}$ 
is related to the minimal area 
of LQG (for more details see 
\cite{Singh:2005xg,Ashtekar:2006wn,Ashtekar:2006uz}). 
Now, using the effective Hamilton equation 
\begin{equation}\label{}
\nonumber
\dot{p}=\{p,\mathcal{H}_{\rm eff}\}=-\frac{\gamma}{3}\frac{\partial
\mathcal{H}_{\rm eff}}{\partial c}
\end{equation}
and the Hamiltonian constraint ($\mathcal{H}_{\rm eff} \approx 0$) 
\cite{Singh:2005xg}
we can obtain the following equations of motion: 
\begin{eqnarray}\label{}
\nonumber
\dot{a}=\frac{1}{\gamma\overline{\mu}}\sin(\overline{\mu} 
c)\cos(\overline{\mu} c)\\
\nonumber \sin^2(\overline{\mu} c)=\frac{8\pi}{3M_p^2
a}\mathcal{H}_m .
\end{eqnarray}
Lastly, combining the above set of equations we provide the first Friedmann 
equation, namely the Hubble parameter 
\begin{equation}
 \label{newfried}
 H^2=\frac{\kappa}{3}\,\rho\,\left(1-\frac{\rho}{\rho_{c}}\right),
\end{equation}
%\begin{equation}\label{g}
%H^{2}\equiv \left(\frac{\dot{a}}{a}\right)^2=\frac{8\pi}{3m_p^2}\rho\left(1-\frac{\rho}{\rho_c}\right),~~~~~~~~\rho_c=\frac{4\sqrt{3}}{\gamma^3} .
%\end{equation}
where %$\kappa=8\pi G = {8 \pi}{M^{-2}_{p}}$, 
$\kappa=8\pi G = M^{-2}_{p}$, 
$\rho$ is the total energy density and 
$\rho_{c}=\sqrt{3}\,/(16\pi^2 G^{2}\gamma^3)\simeq 0.41\rho_{pl},~~\rho_{pl}=M_p^4$ ($M_{p} \simeq 2.43 \times 10^{18}$Gev  is the reduced Planck mass),appears as the critical loop quantum density \cite{Ashtekar:2008gn,Ashtekar:2007em,Zhang:2013yr}.

 Before we continue our analysis we would like to mention 
that in the context of LQC the properties of inflation 
with a standard scalar field 
(the so called ''LQC-inflation'')\footnote{For a standard scalar field 
the corresponding density is given by $\rho_{\phi}=\frac{\dot \phi}{2}+V(\phi)$, 
while in the case of a tachyon field we have 
$\rho_{\phi}=\frac{V(\phi)}{\sqrt{1- {\dot \phi}^2}}$. 
The quantity $V(\phi)$ is the potential.}
has been discussed extensively in the literature
(see \cite{Zhang:2013yr,Copeland:2007qt,Ribassin:2011km,Xiao:2013vaa,Xiong:2007ak,Li:2008tg,Gibbons:2002md,Cai:2010wt,delCampo:2008fc,Setare:2014gya}) 
%(see Refs.[41]-[46] and Ref.[52] CHECK THESE REFERERENCES FROM OUR LIST). 
Here following a similar to the previous papers 
methodology we attempt to investigate the basic features of LQC warm inflation 
with a tachyon scalar field.

%\textbf{ The singularity in general relativity which happens at the begining time of the evolution of the universe, may be replaced by a cosmological bounce in LQC \cite{Ashtekar:2006rx,Ashtekar:2006uz,Singh:2003au, Vereshchagin:2004uc,Bamba:2012ka,Ashtekar:2011rm,Corichi:2010zp}. There are two corrections to the classical dynamic of Friedmann equatiob in LQC, holonomy correction and inverse-volume correction \cite{Ashtekar:2006rx,Ashtekar:2006uz,Singh:2005xg}. We will use holonomy correction of Friedmann equation \eqref{newfried} to study warm-tachyon inflation in the context of LQC. The quadratic density part in Eq.\eqref{newfried}, which is holonomy correction, is important when the scale  of energy density is in order $\rho_c$ \cite{Ashtekar:2008gn, Ashtekar:2007em}.  After big bounce, one of the characters of LQC is super-inflation which naturally happens. In supper-inflation stage where $\frac{1}{2}\rho_c<\rho<\rho_c$, the quantum effect is dominated. Also the Hubble parameter visibly increases and the horizon problem can be solved \cite{Bojowald:2002nz,Bojowald:2003mc,Copeland:2007qt,Ranken:2012hp, Amoros:2013nxa,Xiao:2013vaa}. On the other hand in this epoch $H^2$ is in order $\rho_c$. In this paper we will study super-inflation in the context of warm scenario using tachyon scalar fields.}  

Below, following the work of 
\cite{Xiong:2007ak,Li:2008tg} we introduce the analysis of the tachyon 
field in the framework of LQC.
%In this regime we consider that the cosmic fluid contains a tachyon field
%and radiation which implies that 
%the total density takes the form
%$\rho=\rho_{\phi}+\rho_{\gamma}$, where $\rho_{\phi}$ and $\rho_{\gamma}$
%are the corresponding tachyon field and radiation densities.
Specifically, up to this point we did not specify the nature of the 
fluids involved. Let us now consider that we have a mixture
of two fluids, radiation and tachyon field. Therefore, 
the total density takes the form
$\rho=\rho_{\phi}+\rho_{\gamma}$, where $\rho_{\phi}$ and $\rho_{\gamma}$
are the corresponding tachyon field and radiation densities.
In this framework the overall action \cite{Gibbons:2002md} 
is written as 
$$
S = \int d^{4} x \sqrt{-g} \left(\frac{\cal R}{2\kappa}+L_{\phi}+L_{\gamma}\right) 
$$
where  ${\cal R}$ is the 
Ricci scalar and $g$ is determinant of the metric.

{ At this point it is worth mentioning that 
the nature of tachyon warm inflation has yet to be found. 
However, in order to produce warm inflation one
can use a tachyon scalar field for which the kinetic term
does not follow the canonical form (k-inflation 
\cite{ArmendarizPicon:1999rj}). Therefore, a possible path 
towards understanding the underlying mechanism of 
the current inflationary paradigm is to associate the tachyon fields 
with unstable D-branes \cite{Sen:2002nu}, for which it is well 
known that they lead to 
cosmic acceleration in early times \cite{Sen:2002an, Sami:2002fs, ArmendarizPicon:1999rj}. 
Alternatively, one may study tachyon field inflation 
in the context of the Randall-Sundrum II brane (see Ref.\cite{Setare:2013dd} and references 
therein). 
%VAHID CHECK REFERENCES
}

%It has been presented that tachyonic scalar fields which are associated with
%unstable D-branes \cite{Sen:2002nu} can be responsible 
%for the cosmic acceleration phase in the early times (inflation era)
%As we noticed in introduction, the tachyon potential has the following two properties:
%the maximum of the potential occurs when
%$\phi\rightarrow 0$ while the corresponding minimum of the potential takes place
%when $\phi\rightarrow \infty$. For tachyon inflation models with the ground state at $\phi\rightarrow \infty$, inflaton rolls toward its ground state at infinity without oscillating around  its minimum and the reheating mechanism does not work in this model \cite{Kofman:2002rh}. For quasi power-low time dependence (intermediate), which will be considered in the present work, there is a weak scale factor dependence of the tachyon energy density. Therefore in the post-inflation era the tachyon density would always dominate radiation unless there is a mechanism by which tachyon decay into radiation. There is a solution for reheating problem in the context of warm inflation\cite{ Setare:2012fg, Setare:2013ula, Setare:2014gya,Setare:2014uja,Setare:2013dd,Kamali:2016frd,Kamali:2017zgg,Basilakos:2017bol}, which which will be used in the 
%present work. Warm inflationary scenario can be used for standard inflationary scenarios. For example  MONOMIAL  forms of potential are studied in the context of standard warm inflation \cite{BasteroGil:2009ec}.}    

Notice that
the Lagrangian of the tachyon field \cite{Gibbons:2002md} 
which can be non-minimally coupled to gravity
is given by 
\begin{equation}\label{Lag}
L_{\phi}=
-V(\phi)
\sqrt{1-g^{\mu\nu}\partial_{\mu}\phi\partial_{\nu}\phi} 
\end{equation}
 The radiation Lagrangian $L_{\gamma}$ 
is associated with a fluid of photons with pressure
$P_{\gamma}=\frac{\rho_{\gamma}}{3}$.

Therefore, using the conservation law for the total energy density 
we find 
\begin{equation}\label{Conv}
{\dot \rho}_{\phi}+ 3 H (\rho_{\phi}+p_{\phi}) +{\dot \rho}_{\gamma}+ 4 H\rho_{\gamma} = 0 \;
\end{equation}
 or
\begin{equation}\label{Conv}
{\dot \rho}+ H (3p_{\phi}+3\rho_{\phi}+4 \rho_{\gamma}) = 0 \;,
\end{equation}

This law is the outcome of imposing the covariant 
conservation of the total energy density of the combined 
system of tachyon field and radiation, and thus is a direct 
reflection of the Bianchi identity satisfied by the geometric
side of the Einstein’s equation.
Due to the fact that in warm inflationary scenario \cite{Cai:2010wt,Herrera:2006ck,delCampo:2008fc,Setare:2012fg,Setare:2013dd,Zhang:2013waa}, 
the tachyon-field/photon interaction leads
to radiation production one can split Eq.(\ref{Conv}) as follows
\begin{equation}\label{ss3}
\dot{\rho_{\phi}}+3\,H\,(\rho_{\phi}+p_{\phi})=-\Gamma\dot{\phi}^{2},
\end{equation}
\begin{equation}\label{1.3}
\dot{\rho}_{\gamma}+4H\rho_{\gamma}=\Gamma\dot{\phi}^{2},
\end{equation}
where the positive quantity
$\Gamma$ is the dissipation factor in unit of $M^{5}_{pl}$. 
Usually, in this kind of studies the dissipation term 
$\Gamma \dot{\phi}^2$ is 
given on a phenomenological basis in order to describe the 
nearly-thermal radiation bath of the warm 
inflationary paradigm. Notice, that in several 
papers \cite{Herrera:2006ck, Setare:2012fg, Setare:2013ula, Setare:2014gya, Setare:2014uja, Setare:2013dd} 
one may find another parameter that characterizes warm inflation, namely the 
dimensionless dissipation parameter $R \equiv \frac{\Gamma}{3 H \rho_{\phi}}$.
Within this framework, equation (\ref{ss3}) boils down 
\begin{align}\label{a1}
 \dot \rho_{\phi} = - 3 H \dot \phi^{2} \rho_{\phi}\left(1+R\right).
\end{align}
Notice, that if $R \gg 1$ then we are 
in the high dissipation regime, while 
in the weak dissipation regime
the dimensionless dissipation parameter tends to 
zero ($\Gamma/3H\rho_{\phi} \ll 1$).

Utilizing the energy momentum tensor of the tachyon field
\begin{equation}\label{1.1}
T^{\mu}_{\nu}(\phi)=\frac{\partial
L}{\partial(\partial_{\mu}\phi)}\partial_{\nu}\phi-g^{\mu}_{\nu}L_{\phi}={\rm diag}(-\rho_{\phi},p_{\phi},p_{\phi},p_{\phi})
\end{equation}{equation}
and the corresponding Lagrangian one can obtain 
\begin{equation}\label{1.2}
\rho_{\phi}=\frac{V(\phi)}{\sqrt{1-\dot{\phi}^2}},\ \  p_{\phi}=-V(\phi)\sqrt{1-\dot{\phi}^2}\;.
\end{equation}

 Now differentiating Eq.(\ref{newfried}) with respect to cosmic time $t$ 
and using simultaneously Eq.(\ref{Conv}) we find after some simple calculations
\begin{equation}
 \label{newfried1}
 {\dot H}=-\frac{\kappa}{6}\left(3p_{\phi}+3\rho_{\phi}+4\rho_{\gamma}\right)
\left(1-\frac{2\rho}{\rho_{c}}\right).
\end{equation}
Introducing 
the expressions (\ref{1.2}) into Eq.(\ref{newfried1}) we obtain 
%Now instead of using a canonical scalar field in the above equation 
%(as they have done in aXiv:1311:5325 and references therein), here we 
%are going to introduce a tachyon scalar field, namely
%$\rho_{\phi}=\frac{V(\phi)}{\sqrt{1- {\dot \phi}^2}}$ and 
%$p_{\phi}=-V(\phi)\sqrt{1- {\dot \phi}^2}$. Inserting the latter 
%expressions 
\begin{equation}
 \label{newfried2}
 {\dot H}=-\frac{\kappa}{6}\left(3\rho_{\phi}{\dot \phi}^{2}+4\rho_{\gamma}\right)
\left(1-\frac{2\rho}{\rho_{c}}\right).
\end{equation}
Obviously, in order to understand the dynamical behaviour 
of the current problem we need to combine 
Eq.(\ref{newfried}) with Eq.(\ref{newfried2}). 
In particular, from Eq.(\ref{newfried}) the quantum bounce point ($H=0$) 
occurs for $\rho=\rho_{c}$ \cite{Ashtekar:2006rx,Ashtekar:2006uz,Singh:2003au, Vereshchagin:2004uc,Bamba:2012ka,Ashtekar:2011rm,Corichi:2010zp,Singh:2005xg,Bojowald:2002nz,Bojowald:2003mc,Copeland:2007qt,Ranken:2012hp, Amoros:2013nxa,Xiao:2013vaa}. On the other hand, prior to the 
bounce point the quantity ${\dot H}$ satisfies the restriction
${\dot H}>0$ and it remains in the positive regime until 
$\rho=\frac{\rho_{c}}{2}$. At this point ${\dot H}$ vanishes and eventually 
it becomes negative. 
Following the notations Ref.\cite{Xiao:2013vaa}, we verify 
%accomodates a ''super-tachyon inflation''
%From the aforementioned analysis it becomes clear 
that LQC accommodates a ''LQC-tachyon inflation''
from $\rho=\rho_{c}$ to 
$\rho=\frac{\rho_{c}}{2}$ ($\frac{\rho_{c}}{2}< \rho < \rho_{c}$). 
In this regime the Hubble parameter lies in the interval 
$0<H< H_{I}$, where the characteristic scale 
$H_{I}=\left(\frac{\kappa \rho_{c}}{12}\right)^{1/2}$ corresponds to 
the epoch where $\rho=\frac{\rho_{c}}{2}$ and for which 
''LQC-tachyon inflation'' ends.
{ It is interesting to mention that 
the same to the above condition 
(see $\frac{\rho_{c}}{2}< \rho < \rho_{c}$) holds 
in the case of
%it has been found that 
''LQC-inflation''\cite{Zhang:2013yr,Copeland:2007qt,Ribassin:2011km,Xiao:2013vaa,Xiong:2007ak,Li:2008tg,Gibbons:2002md,Cai:2010wt,delCampo:2008fc,Setare:2014gya} in which the scalar field has 
the standard form ($\rho_{\phi}=\frac{\dot \phi}{2}+V$).}
%This inflationary paradigm 
%lives in the quantum geometry dominated era.
%Therefore, during the LQC warm tachyon inflation the quantity 
%$\frac{H^{2}}{\kappa \rho_{c}}$ is not negligible.
To this end, the arofementioned analysis points 
%Eq.(\ref{newfried2})
that { the $(1-\rho/\rho_{c})$ term, imposes a specific 
era of inflation $0<H<H_{I}$}, where 
for $\rho \simeq \rho_{c}$ the universe enters in the  
inflationary phase, while the characteristic scale $H_{I}$ 
presumably connected to a scale which is around 
to that of the Grand Unified Theory (GUT). 
For example, substituting 
$\kappa= M^{-2}_{p}$, $M_{p}\simeq 2.43\times 10^{18}$Gev and
$\rho_{c} \simeq 0.41M_{p}^{4}$ into 
$H_{I}=\left(\frac{\kappa \rho_{c}}{12}\right)^{1/2}$ we find 
$H_{I}\sim 10^{17}$ Gev.
Well after the primeval inflationary epoch, specifically for 
$\rho \ll \rho_{c}$, the Universe enters in 
the nominal radiation era. After this period the radiation 
component starts to become sub-dominant and the matter dominated era appears.

{% It was shown in Ref.\cite{Xiao:2013vaa}: There is small e-folding number during super-inflation stage expect the potential dominated at the bounce point $\rho\sim V(\phi)$. At this point $H=0$ and $\dot{H}$ is positive, immediately after this point the universe enters a slow-roll super-inflation (with condition $\rho\sim V(\phi)$). During this epoch the kinetic energy of the scalar field is increasing and at the same time this energy transfers to radiation energy density (warm inflation). We note that the Hubble parameter is not constant in this stage \cite{Copeland:2007qt} (We will use intermediate inflation with this condition in our study) which can be used to solve the horizon problem. Solving the horizon problem, $aH$ must grow sufficiently during super inflation era. $\bar{N}\equiv \ln\frac{a(t_f)H(t_f)}{a(t_i)H(t_i)}$ has to be $60$ at least \cite{Amoros:2013nxa}. For standard inflation $H\simeq const$  and $\bar{N}=N\equiv\ln\frac{a_f}{a_i}$. If the potential is dominating at the bounce point, the super-inflation can support enough e-folding number \cite{Xiao:2013vaa}. 
%}
{ Overall, %for both LQC inflationary models 
the condition
$H=0$, ${\dot H}>0$ implies that 
the Universe enters in the %super-inflation 
LQC inflationary era at the bounce point
\cite{Xiao:2013vaa} (see also \cite{Xiong:2007ak}). 
%Since the LQC-warm tachyon inflation
%shares exactly the same dynamical condition 
%with the LQC-inflation means 
%that the former inflationary scenario inherits all the merits and demerits
%of the latter model.
%Particular attention over the last decade has been
%paid on the  horizon problem in the LQC-inflation 
%(''super-inflation''; see Refs. \refcite{Copeland:2007qt,Ribassin:2011km,Xiao:2013vaa,Xiong:2007ak,Ranken:2012hp,Amoros:2013nxa}).
%[37],[39], [58], [59]).
Unlike nominal (cold) inflation, in which $H\simeq $const., here 
the key point is to understand that the Hubble parameter varies with time.
Therefore, it has been proposed (see Ref.\cite{Xiao:2013vaa} 
references therein) 
that the necessary condition 
towards solving the horizon problem in the context of 
%super
LQC-inflation is to ensure that the quantity $aH$ grows 
substantially during the early period of the Universe evolution.
Specifically, it has been shown \cite{Copeland:2007qt,Ribassin:2011km,Xiao:2013vaa} that the horizon problem 
can be solved if one introduces a single scalar field 
(for other alternatives see \cite{Ranken:2012hp}) in 
%super
LQC-inflation. Also, Ref.\cite{Amoros:2013nxa} found that %super
LQC-inflation 
can provide the appropriate number of e-folds ($\sim 60$), while 
the Universe includes a small cosmological constant and  
matter.}

Now, inserting the equations (\ref{1.2}) 
into Eq.(\ref{ss3}) we 
immediately derive the modified Klein-Gordon equation 
which provides the time-evolution of $\phi$
\begin{align}\label{cv1}
\frac{\ddot \phi}{1-\dot \phi^{2}} + 3 H \dot \phi + \frac{1}{V} 
\frac{dV}{d\phi} 
= - \frac{\Gamma}{V} \dot \phi \sqrt{1-\dot \phi^{2}}.
\end{align}
In the slow-roll approximation $\dot{\phi}^2\ll 1$ for tachyon field, the above equation simplified as:
 \begin{align}\label{cv2}
\ddot \phi + (3H+\frac{\Gamma}{V}) \dot \phi + \frac{1}{V} 
\frac{dV}{d\phi} 
= 0 .
\end{align}
As it is obvious from the above equation the energy exchange 
between the tachyon field and radiation introduces 
an additional friction term ($\frac{\Gamma}{V}\dot{\phi}$) that modifies 
the standard Klein-Gordon equation. 

%The point can readily be deduced observing Eq. (\ref{cv1}) is 
%that the energy exchange between inflaton field and other 
%quantum fields during inflationary phase in the leading 
%adiabatic approxiamation modifies the dynamical equation 
%of inflation by a supplementary friction term. The 
%immediate consequence of such additional friction term is 
%that steeper potential can indeed provide appropriate 
%e-folding number to overcome aforementioned shortcomings 
%in introduction. 

At the epoch of warm inflation 
it is safe to assume that 
the energy density of the tachyon field
is the dominant component of the cosmic fluid ($\rho_{\phi}\gg \rho_{\gamma}$ 
see \cite{BasteroGil:2012zr}). Therefore,  
the effective Friedmann Eq. (\ref{newfried}) reduces to
\begin{align}\label{sp5}
 H^{2} = \frac{\kappa}{3} \rho_{\phi} \left(1  -\frac{\rho_{\phi}}{\rho_{c}}\right).
 \end{align}
On the other hand, 
utilizing Eqs. (\ref{a1}, \ref{sp5}), one can easily show that
\begin{align}
\label{dotphi.1}
 \dot \phi^{2} &= - \frac{2\dot H }{ \kappa  \rho_{\phi} (1+R)} \left(1 - \frac{12 H^{2}}{\kappa \rho_{c}}\right)^{-\frac{1}{2}}.
\end{align}
Moreover, if we consider that the quantity 
$\Gamma \dot \phi^{2}$ varies adiabatically then 
the radiation component evolves slowly which means that 
$\dot{\rho}_{\gamma}\ll4H\rho_{\gamma}$ and 
$\dot{\rho }_{\gamma}\ll\Gamma\dot{\phi}^{2}$. 
Under the latter conditions the combination of 
Eq. (\ref{1.3}) and Eq. (\ref{dotphi.1}) yields  
\begin{equation}
\label{sp55}
\rho_{\gamma} = \frac{\Gamma\dot{\phi}^{2}}{4H}=\frac{- \Gamma \dot{H}}{2 \kappa
H(1+R)\rho_{\phi}} \left(1 - \frac{12 H^{2}}{\kappa \rho_{c}}\right)^{-\frac{1}{2}}.
\end{equation} 
On the other hand, 
under adiabatic condition the above formula 
can be identified with the expression relating
$\rho_{\gamma}$ with the radiation temperature $T$.
By adding all the degrees of freedom of the created massless modes, 
the relationship between the radiation energy density and 
the temperature is given by
\cite{Kolb:1990vq}, 
\begin{align}
\rho_\gamma=C_\gamma T^4
\end{align}
and combining with Eq.(\ref{sp55}) we find 
\begin{align}\label{tem}
T=\left[  \frac{-\Gamma\dot{H}}{2 \kappa  C_{\gamma} H(1+R)\rho_{\phi}} \right]^{\frac{1}{4}}\left(1 - \frac{12 H^{2}}{\kappa \rho_{c}}\right)^{-\frac{1}{8}},
\end{align}
where $C_\gamma=\pi^2 g_*/30$ in which we included 
the degrees of freedom of the created massless modes
via the $g_*$ factor.
Substituting Eq. (\ref{tem}) in (\ref{G}) we obtain  
\begin{align}\label{gam}
 \Gamma^{1-\frac m4}&\left(1+R\right)^{\frac m4} = C_{\phi} \phi^{1-m} \left( \frac{-\dot{H}}{2 \kappa  C_{\gamma}
H\rho_{\phi}}\right)^{\frac{m}{4}}\left(1 - \frac{12 H^{2}}{\kappa \rho_{c}}\right)^{-\frac m8}.
\end{align}
Now, by solving Eq.({\ref{sp5}) 
for the tachyon density and using simultaneously % ({\bf CHECK CALCULATIONS}) 
Eq.(\ref{dotphi.1}) we find:
\begin{align}\label{pot}
\nonumber V(t)&= \frac{\rho_{c}}2\left({1- \sqrt{1 - \frac{12H^{2}}{\kappa \rho_{c}}}}\right) \\& \times {\left[1+\frac{2\dot H }{ \kappa  \rho_{\phi} (1+R)} \left(1 - \frac{12 H^{2}}{\kappa \rho_{c}}\right)^{-\frac{1}{2}} \right]^{\frac12}}.
\end{align}
Lastly, including the slow-roll approximation $\dot \phi^{2}\ll 1$ in 
Eq.(\ref{1.2}) we have $\rho_{\phi} \simeq V(\phi)$ and thus 
$R = \frac{\Gamma}{3 H V}$. Therefore, Eq. (\ref{pot}) reduces to
\begin{align}\label{qw1}
V(t)\simeq \frac{\rho_{c}}2\left({1- \sqrt{1 - \frac{12H^{2}}{\kappa \rho_{c}}}}\right) .
\end{align}

%\textbf{ In Eq.(\ref{qw1}), term $- \frac{12 H^{2}}{\kappa \rho_{c}}$ in R.H.S is warm-LQC (warm super-inflation) modification which is smaller than one but in a same order of one $\mathcal{O}(1)$ as we discussed after Eq.(\ref{newfried}) .
%As expected, the potential energy 
%is given in terms of the  scale of the loop quantum density $\rho_{c}$.}

From the above analysis it becomes clear that the aforementioned 
cosmological functions strongly dependent on 
the dissipation factor $\Gamma$ 
and the background expansion history, through the Hubble parameter $H(t)$.
As we have already mentioned 
the tachyon field is exchanging energy with 
radiation implying that the dissipation factor
is a characteristic quantity of warm inflation. 
Although, the precise form of the dissipation factor 
has yet to be found,
various candidates for $\Gamma$ have been proposed in the literature. 
In the present study, based on the notations 
\cite{BasteroGil:2011xd,Zhang:2009ge} we treat the dissipation factor
as follows 
\begin{equation}
\Gamma(\phi, T)=C_{\phi}\,\frac{T^{m}}{\phi^{m-1}}, \label{G}
\end{equation}
where $T$ is the temperature, $C_{\phi}$ is constant. 
%and 
%$m$ is an exponet 
%which determines the form of $\Gamma$. 
From now on, we will call $\Gamma_{m}$ the parametrization 
of the dissipation factor
where the index $m$ determines the form of $\Gamma$. 
%is related with the undelying physics
%of the tachyon-field/photon interaction.
Under the latter parametrization 
we immediately recognize the following situations: (a) 
when $m=-1$ (hereafter $\Gamma_{-1}$) 
we have $\Gamma = C_{\phi} \frac{\phi^{2}}{T}$
which corresponds to the dissipation rate of the 
non-SUSY model \cite{Berera:1998gx,Yokoyama:1998ju,Herrera:2013rra}, 
(b) for $m=0$ ($\Gamma_{0}$ parametrization)  
we get $\Gamma = C_{\phi} \phi$ 
(SUSY case see \cite{Berera:1998gx})
(c) for $m=1$ ($\Gamma_{1}$ parametrization) we have 
$\Gamma = C_{\phi} T$ (see Refs.\cite{Panotopoulos:2015qwa,Berera:2008ar,Moss:2008yb,Bastero-Gil:2016qru})
and (d) for $m=3$ ($\Gamma_{3}$ parametrization) we obtain  
$\Gamma = C_{\phi} \frac{T^{3}}{\phi^{2}}$
that corresponds to the low temperature SUSY model
\cite{Moss:2006gt,BuenoSanchez:2008nc,Ramos:2013nsa,Cerezo:2012ub}.

In this kind of studies it is well known that 
the precise functional form of the 
potential $V(t)$ and the Hubble parameter $H(t)$
can not be simultaneously found from first principles. The latter 
implies that the only way
to use the Loop Quantum approach in Cosmology is to
phenomenologically select the functional form 
of either the potential or the Hubble parameter. 
As an example in the context of LQC with a standard scalar field 
($\rho_{\phi}=\frac{\dot \phi}{2}+V$) we refer the reader the work 
of Ref.\cite{Zhang:2013yr} %(CHECK REFERENCE). 
These authors have been phenomenologically introduced 
the well known power law 
potential $V \propto  \phi^{2}$ in LQC towards treating 
the cosmic expansion. Alternatively, one may 
impose the functional form of the Hubble parameter 
in order to derive the potential \cite{Barrow:1990td,Barrow:1993zq,Berera:1996fm,Setare:2012fg,Setare:2013ula,Herrera:2013rra,Sharif:2015oba}.
%Clearly, each approach has advantages and disadvantages, but it is important
%to realize that for both treatments 
%the evolution of the basic cosmological functions 
%(Hubble parameter and potential) are affected  
%by the critical loop quantum density since the cosmic time 
%$t$ can be written in terms 
%of $\rho_{c}$ [cf. Eq.(18) in arXiv:1311.5325 and 
%Eqs. (27)-(28) in our paper].

In our work we have decided 
to phenomenologically select the functional form of $H(t)$ using the well
known solution of warm intermediate inflation provided by 
Barrow\cite{Barrow:1996bd} as a reference model.

%Regarding the functional 
%form of the Hubble parameter $H(t)$ we utilize  
%Barrow's \cite{Barrow:1996bd} solution for 
%the scale factor of the universe. 
%In particular, 
{ Inspired by the work of Herrera et al.\cite{Herrera:2014mca} 
in the context of warm intermediate inflation, we consider the following 
form of  
the scale factor}% obeys the following exponential expression:
\begin{eqnarray}\label{scall}
a(t)=a_{I}\exp(At^f)\;, 
\end{eqnarray}
where $f$ lies in the interval $0<f<1$.
Obviously, the evolution of Hubble parameter and 
its first time-derivative are given by 
\begin{align}\label{sd1}
H = \frac{\dot a}{a} = A f t^{f-1}, \ \ \ \  
\dot H = A f \left(f-1\right) t^{f-2}.
\end{align}
In this scenario the cosmic expansion evolves 
slower than the standard de Sitter model, $a(t)\propto {\rm exp}(H_{I}t)$
[$H(t)=H_{I}=$const.] and 
faster than the power-law inflation
($a \propto t^{p}$, $p>1$).
{ The exponential expression of the scale factor Eq.(\ref{scall}) 
and the corresponding Hubble parameter
Eq.(\ref{sd1}) are used in order to approximate the cosmic expansion 
prior to the inflationary era and not near the bounce.
Therefore, Eq.(\ref{sd1}) is obtained using the classical Einstein's
equations. Similar to our notations have been used in 
the paper of Herrera et al.\cite{Herrera:2014mca} but in the 
case of LQC warm inflation with 
a canonical scalar field.}

%Q3: The above form of the scale factor (\ref{scall}) 
%and Hubble evolution (\ref{sd1}) 
%is obtained for inflationary phase using classical Einstein's equations. This form of Hubble parameter is exactly used in inflation era not near the bounce. There is an approximation that we will use LQC effect  in the inflation era 
%\cite{Xiao:2014nda,Herrera:2014mca}.
%{\bf CHECK In figure 5 we plot the predicted
%$(n_s, r)$ in the case non-LQC warm inflation (see stars in figure 5). We then compare the
%latter $(n_s, r)$ predictions with those of the LQC warm tachyon inflation (solid points). In
%principle, this can help us to understand better the theoretical expectations of the LQC
%warm tachyon inflationary model, as well as to identify the differences from the non-LQC warm tachyon inflation.} 

%%%%%\textbf{ so the name of this evolution of scale factor is "intermediate" \cite{Barrow:1996bd}}.

\section{The Strong Dissipative Regime and Slow-Roll parameters:}
%\section{The Strong Dissipative Regime and Slow-Roll parameters}
In the rest of the paper we focus our analysis 
on the high dissipation regime ($R\gg1$). 
Hence, Eqs. (\ref{dotphi.1}, \ref{gam}) reduces to
\begin{align}
 \dot \phi^{2} &= - \frac{6 H \dot H }{ \kappa \Gamma} \left(1 - \frac{12 H^{2}}{\kappa \rho_{c}}\right)^{-\frac{1}{2}},\label{sm1}
\end{align}

\begin{equation}\label{2.1}
\Gamma=C_{\phi} \phi^{1-m} \left(-\frac32 \frac{\dot H}{C_\gamma \kappa}\right)^{\frac m4}\left(1-\frac{12 H^2}{\kappa \rho_c}\right)^{-\frac m8}.
\end{equation}
If we insert Eq. (\ref{2.1}) in 
Eq. (\ref{sm1}) then 
the evolution of tachyon scalar field is given by
\begin{align}\label{dotphi2}
 \dot\phi=\left[\frac{4 C_{\gamma}}{C_{\phi}} H \phi^{m-1} \left(- \frac{3 \dot H }{2\kappa C_{\gamma}}\right)^{\frac{4-m}{4}} \left(1 - \frac{12H^{2}}{\kappa \rho_{c}}\right)^{\frac{m-4}{8}} \right]^{\frac{1}{2}}.
\end{align}
Performing the integration of Eq. (\ref{dotphi2}) with the aid of 
Eq. (\ref{sd1}) we have 
\begin{equation}
\label{fft}
\phi(t)-\phi_{\star}=\left\{ \begin{array}{cc}
       \left[\frac{(3-m)G_{m}(t)}{2K_{m}}\right]^{2/(3-m)} &
       \mbox{for $m\ne 3$}\\
       {\rm exp}\left[\frac{G_{m}(t)}{K_{m}}\right] & \mbox{for $m=3$}
       \end{array}
        \right.
\end{equation}
{ where $\phi_{\star}$ is the value of $\phi(t_{\star})$ 
at a characteristic time $t_{\star}$ which obeys the 
following inequality $t_{\star} >t_{\rm bounce}$ by definition.} 
%at the time $t_0$ which we start integration on $\dot{\phi}$ Eq.(\ref{dotphi2}). This time is in the inflation period. }
In the above relations  $K_{m}$ and $G_{m}(t)$ are defined as follows
\begin{align}\label{ff1}
\nonumber K_{m} &\equiv \frac{(8-m)f+2m-4}{8}\left(\frac{4 A f C_{\gamma}}{C_{\phi}}\right)^{-\frac{1}{2}}\\ & \times \left(\frac{3 A f (1-f)}{2 \kappa C_{\gamma}}\right)^{\frac{m-4}{8}}.
\end{align}
\begin{align}\label{ff2}
\nonumber G_{m}(t) &\equiv t^{\frac{(8-m)f +2m-4}{8}} {}_1 F_{2}\left[\frac{4-m}{16}, \frac{(8-m)f +2m-4}{16(f-1)}\right. \\& \left. ,1+ \frac{(8-m)f +2m-4}{16(f-1)}, S{t^{2f-2}}\right].
\end{align}
Notice, that ${}_1 F_{2}$ is the hypergeometric function 
and $S \equiv \frac{12 A^{2} f^{2}}{\kappa \rho_{c}}$. 
Without any loss of generality we set $\phi_{\star}=0$.

Under the above conditions the Hubble parameter 
can be expressed as a function of the tachyon scalar field as follows

\begin{align}
\label{wr1}
H(\varphi) &= A f \left(G_{m}^{-1}\left(K_{m} \varphi \right)\right)^{f-1},
\end{align} 
in which $G_{m}^{-1}$ corresponds to the inverse function of 
$G_{m}(t)$. Moreover, the variable $\varphi$ is given by
\begin{equation}
\label{fft1}
\varphi=\left\{ \begin{array}{cc}
       \frac{2\phi^{(3-m)/2}}{3-m} &
       \mbox{for $m\ne 3$}\\
       {\rm ln}\left(\phi \right) & \mbox{for $m=3$.}
       \end{array}
        \right.
\end{equation}

 Clearly, the evolution of the Hubble parameter is affected  
by the critical loop quantum density since the cosmic time 
$t$ can be written in terms of $\rho_{c}$ [see Eqs. (\ref{ff2}) and
(\ref{wr1})].
Therefore, using Eqs.(\ref{qw1}, \ref{sd1}, \ref{2.1}, \ref{ff2}),%\ref{p1}, \ref{p2}), 
the potential and the dissipation factor 
are calculated as:
\begin{align}
V(\varphi) &\simeq \frac{\rho_{c}}{2} \left[1 -  \left(1 - S \left(G_{m}^{-1}(K_{m}\varphi)\right)^{2f-2}\right)^{\frac{1}{2}}\right]~~,\label{wr3}
\end{align}
and 
\begin{align}
 \nonumber \Gamma (\varphi)&= C_{\phi} \varphi^{1-m} \left( \frac{3 A f (1-f)\left(G^{-1}_{m}({K}_{m}\varphi\right)^{f-2}}{2 \kappa C_{\gamma} } \right)^{\frac{m}{4}}\\ & \times \left(1 - S \left(G^{-1}_{m}({K}_{m}\varphi)\right)^{2f-2}\right)^{-\frac{m}{8}}.\label{wr5}
\end{align}

Now, using standard lines we are ready to provide the slow-roll parameters
\begin{align}
\epsilon \equiv - \frac{\dot H}{H^{2}}, \ \ \ \ \eta \equiv - \frac{\ddot H}{2 H \dot H}.
\end{align}
Indeed, with the aid of Eqs. (\ref{sd1},\ref{2.1}, \ref{ff2}% \ref{p1}, \ref{p2}
) the slow-roll parameters 
become
\begin{align}
\epsilon & =\frac{1-f}{A f t^{f}}= \frac{1-f}{A f \left(G^{-1}_{m}(K_{m} \varphi)\right)^{f}},\label{sg1}\\
%\epsilon & = \frac{1-f}{A f \left(G^{-1}_{3}(K_{3} \ln \phi)\right)^{f}}, \label{sg2}\\
\eta &=\frac{2-f}{2 A f t^{f}}= \frac{2 -f}{2 A f \left(G^{-1}_{m}(K_{m}\varphi)\right)^{f}}.\\
%\eta &=\frac{2 -f}{2 A f \left(G^{-1}_{3}(K_{3}\ln \phi)\right)^{f}}.
\label{wr6}
\end{align}
We would like to stress here that for the intermediate inflation 
the slow-roll parameters are always less than unity which means that 
inflation never ends. In our model the condition 
$\epsilon=1$ makes sure that inflation starts at the earliest possible
stage \cite{Barrow:2006dh, Barrow:1993zq}. 
As far as the number of e-folds is concerned we have 
\begin{align}\label{cc1}
N=\int^{t_{k}}_{t_{in}} H dt=A \left(t_{k}^{f} - t_{in}^{f}\right)
\end{align}
or
\begin{align}
N&=  A \left(\left(G^{-1}_{m}(K_{m}\varphi_{k})\right)^{f} -\left(G^{-1}_{m}(K_{m}\varphi_{in})\right) ^{f}\right)\label{mm1},
\end{align}
where for the last equality we used Eq.(\ref{sd1}). 
Also, $\varphi_{k}=\varphi(t_{k})$ and $\varphi_{in}=\varphi(t_{in})$ denote
the values at the horizon crossing 
and at the beginning of inflation respectively. 
Equating Eqs. (\ref{sg1}) to unity, $\epsilon \left(\varphi_{in}\right) =1$ 
and using Eqs.(\ref{fft}) and (\ref{fft1}) 
the value of the tachyon scalar field at the beginning of inflation is given by
\begin{equation}
\phi_{in}=\left\{ \begin{array}{cc}
       [\frac{(3-m)G_{m}(y)}{2K_{m}}]^{2/(3-m)} &
       \mbox{for $m\ne 3$}\\
       {\rm exp}\left[\frac{G_{m}(y)}{K_{m}}\right] & \mbox{for $m=3$}
       \end{array}
        \right.
        \label{wr7},        
\end{equation}
where $y=\left(\frac{1-f}{Af}\right)^{1/f}$.
To this end, it becomes clear that from (\ref{wr7}) one may 
express Eqs. (\ref{wr1} - \ref{wr6}) as a function of $N$. Indeed, by doing
that we find 
\begin{equation}
\phi_{k}=\left\{ \begin{array}{cc}
       \left[\frac{(3-m)G_{m}(I(N))}{2K_{m}}\right]^{2/(3-m)} &
       \mbox{for $m\ne 3$}\\
       {\rm exp}\left[\frac{G_{m}(I(N))}{K_{m}}\right] & \mbox{for $m=3$}
       \end{array}
        \right.
\end{equation}
where $I(N)=(\frac{1+f(N-1)}{fA})^{\frac{1}{f}}$.
One of the most important features of any inflationary scenario 
is related with the formation of large scale structures. 
For example, in the case of warm inflationary paradigm 
thermal fluctuations play a key role because they provide 
the initial seeds for the formation of cosmic structures.   
The situation regarding  
cosmological perturbations within the effective Hamiltonian formalism 
in LQC has been studied in \cite{Bojowald:2006tm, WilsonEwing:2012bx,Bojowald:2008gz,Cailleteau:2011kr}, which however is beyond the scope of the present 
study.
%The quantum effect is weak when Hubble crossing 
%as stated in \cite{Zhang:2007bi,Zhang:2009ge,Herrera:2010yg}, therefore, 
 Cosmological perturbations in LQC have been explored  
in several studies 
\cite{Zhang:2009ge,Herrera:2010yg,Haro:2013bea,Setare:2014gya,Zhang:2007bi,Herrera:2010yg}. 
%Unlike standard cold inflation, here the scalar perturbations are thermal 
%field is allowed to interact with other light fields, 
Following these works the corresponding curvature perturbation
was found to be $\delta {\cal R}=\left(\frac{H}{\dot \phi} \right) \delta \phi$ 
and thus the amplitude of scalar fluctuations for LQC is given by
%${\cal P}_{\cal R}=\frac{H^2}{{\dot \phi}^2} \delta \phi^{2}$. 

%The scalar perturbation $\delta\phi$ in cold scenario 
%may be treated as perturbation of a massless scalar field \cite{Zhang:2009ge} 
%which is proportional to Hubble parameter, but in warm scenario 
%thermal perturbation $\delta\phi$ is proportional to temperature of thermal bath and the curvature perturbation is presented by $R\simeq\frac{H}{\dot{\phi}}\delta\phi$. } 

%Following Refs.\cite{Berera:1995ie,Zhang:2007bi,Zhang:2009ge,Herrera:2010yg} 
%the corresponding power spectrum of scalar fluctuations 
%for LQC is presented by
\begin{align}\label{se2}
\mathcal{P}_{\mathcal{R}} = \frac{H^{2}}{\dot \phi^{2}} \delta \phi^{2}.
\end{align}
As we have already defined in the introduction 
the nature of scalar perturbations in warm inflation is 
thermal and not quantum as
we consider in the standard inflationary model.
In particular, 
it has been found \cite{Berera:1995ie, Berera:1996fm, Hall:2003zp}
that in the case of $R \gg 1$
warm scalar perturbations obey the following expression 
\begin{eqnarray}\label{2.12}
\delta\phi^2\simeq\frac{k_F T}{2 M_{pl}^4\pi^2} \;,
\end{eqnarray}
where the wave number 
$k_F=\sqrt{\frac{\Gamma H}{V}}=H\sqrt{\frac{\Gamma}{3HV}}\geq H$ 
provides the freeze-out scale at which 
the dissipation damps out to thermally excited 
fluctuations of inflaton 
($\frac{V''}{V'}<\frac{\Gamma H}{V}$) \cite{Taylor:2000ze}. 
Inserting the wave-number $k_{F}$ and Eq.(\ref{2.12}) into 
Eq.(\ref{se2}) we arrive at 

\begin{align}\label{pow}
\mathcal{P}_{\mathcal{R}} = \frac{\kappa^2 H^{\frac{5}{2}} \Gamma^{\frac{1}{2}} T}{128 \pi^{4}  V^{\frac{1}{2}} \dot \phi^{2}}.
\end{align}
We continue our calculations  by 
substituting Eqs. (\ref{wr1} - \ref{wr6}) 
and Eqs. (\ref{wr7}) in Eq. (\ref{pow}) 
we obtain the power 
spectrum in terms of the number of e-folds

\begin{align}
\nonumber \mathcal{P}_{\mathcal{R}} &= P_{m }\left(I(N)\right)^{\frac{3m+6}{8}f - \frac{3m}{4}} \left(1- S \left(I(N)\right)^{2f-2}\right)^{\frac{6-3m}{16}} \\& \times \left(1- \left(1- S \left(I(N)\right)^{2f-2}\right)^{\frac{1}{2}}\right)^{-\frac{1}{2}} \left(\frac{G_{m}(I(N))}{{K}_{m}}\right)^{\frac{3(1-m)}{2}},
\end{align}
%\begin{align}
%\nonumber \mathcal{P}_{\mathcal{R}} &= P_3 \left(I(N)\right)^{\frac{15}{8}f - \frac{9}{4}} \left(1- S \left(I(N)\right)^{2f-2}\right)^{-\frac{3}{16}} \\ & \times\left(1- \left(1- S \left(I(N)\right)^{2f-2}\right)^{\frac{1}{2}}\right)^{-\frac{1}{2}} \exp \left(- \frac{3 G_{3}\left(I(N)\right)}{K_{3}}\right).
%\end{align}
where
\begin{align}
P_{m} &= \left(\frac{\kappa^{2} (Af)^{\frac{3}{2}} C_{\phi}^{\frac{3}{2}}}{256 \sqrt{2} \pi^{4} \sqrt{\rho_{c}} C_{\gamma}}\right) \left( \frac{3 A f (1-f) }{2 \kappa C_{\gamma} } \right)^{\frac{3m-6}{8}}. 
%\\
%P_3 &= \left(\frac{\kappa^2 (Af)^{\frac{3}{2}} C_{\phi}^{\frac{3}{2}}}{256 \sqrt{2}\pi^{4} \rho_{c} C_{\gamma}}\right) \left( \frac{3 A f (1-f)}{2\kappa C_{\gamma} } \right)^{\frac{3}{8}}.
\end{align}
At this point we introduce the spectral spectral index in our analysis
which is defined as

\begin{align}
n_{s} -1 \equiv \frac{d \ln \mathcal{P}_{\mathcal{R}}}{d \ln k}.
\end{align}
Since $d{\rm ln}k=-dN$ and  armed with 
the analytic expression of 
${\cal P}_{\cal R}$ [see Eq.(\ref{se2})] we derive 
%$n_{s} -1 \equiv \frac{d \ln \mathcal{P}_{\mathcal{R}}}{d \ln k}$
%we can define the spectral index of scalar perturbations (see equations 46-49).
after some simple algebra the scalar spectral index takes 
%the form
\begin{align}\label{sz1}
n_{s}-1 &= {n}_{1}+ {n}_{2}+{n}_{3}+ {n}_{4},
\end{align}
with
\begin{align}\label{sz1v}
{n}_{1}& = -\frac{{\frac{3m+6}{8}f - \frac{3}{4}}}{A f \left(I(N)\right)^{f}},\\
{n}_{2}& =  -\frac{(6-3m) S (f-1)\left(I(N)\right)^{f-2}}{8Af \left(1- S\left(I(N)\right)^{2f-2} \right)},\\
{n}_{3}&=  \frac{(f-1) S \left(I(N)\right)^{f-2} \left(1- S \left(I(N)\right)^{2f-2}\right)^{-\frac{1}{2}}}{2 A f  \left(1- \left(1- S \left(I(N)\right)^{2f-2}\right)^{\frac{1}{2}}\right) },\\
\nonumber {n}_{4}& =  -\frac{3(1-m) \left((8-m) f+ 2m-4\right)}{8 K_{m}} \left(I(N)\right)^{-\frac{m}{8}(f-2)- \frac{1}{2}}
\end{align}
\begin{align}
\times \left(1- S \left(I(N)\right)^{2f-2}\right)^{\frac{m-4}{16}} \left(2A f \left(\frac{G_{m}(I(N))}{K_{m}}\right)^{\frac{3-m}{2}}\right)^{-1}.
\end{align}

On the other hand, the amplitude of tensor fluctuations  
is given by 
\begin{align}
\mathcal{P}_{t} = 8 \kappa \left(\frac{H}{2 \pi}\right)^{2},
\end{align}
from which we can define the tensor-to-scalar ratio

\begin{align}\label{vv2}
r \equiv \frac{\mathcal{P}_{t}}{\mathcal{P}_{\mathcal{R}}} =  \frac{256 \pi^{2} V^{\frac{1}{2}} \dot \phi^{2}}{\kappa H^{\frac{1}{2}} \Gamma^{\frac{1}{2}} T} . 
\end{align}
Introducing the appropriate formulas of $H$, $V$, $\Gamma$ and $T$   
in the above equation we find 
\begin{align}
\label{sz11}
\nonumber r &= r_{m} \left(I(N)\right)^{\frac{10 - 3m}{8}f + \frac{6m -16}{8}} \left(1- S \left(I(N)\right)^{2f-2}\right)^{\frac{3m -6}{16}} \\& \times \left(1- \left(1- S \left(I(N)\right)^{2f-2}\right)^{\frac{1}{2}}\right)^{\frac{1}{2}} \left(\frac{G_{m}(I(N))}{{K}_{m}}\right)^{\frac{3(1-m)}{2}}
\end{align}
%\begin{align}
%\nonumber r & = r_{3} \left(I(N)\right)^{\frac{1}{8}f +\frac{1}{4}} \left(1- S \left(I(N)\right)^{2f-2}\right)^{\frac{3}{16}} \\& \times\left(1- \left(1- S \left(I(N)\right)^{2f-2}\right)^{\frac{1}{2}}\right)^{\frac{1}{2}} \exp \left(- \frac{3 G_{3}\left(I(N)\right)}{{K}_{3}}\right).
%\end{align}
where
\begin{align}
r_{m} &= \frac{256 \pi^{2}C_{\gamma}}{\kappa C_{\phi}^{\frac{3}{2}}} \sqrt{{2 \rho_{c}  A f}}\left( \frac{3 A f (1-f) }{2\kappa C_{\gamma} } \right)^{\frac{6-3m}{8}}.%,\\
%r_{3} &= \frac{256 \pi^2 C_{\gamma}}{\kappa C_{\phi}^{\frac{3}{2}}} \sqrt{{2 \rho_{c}  A f}}\left( \frac{3 A f (1-f) }{2\kappa C_{\gamma} } \right)^{-\frac{3}{8}}.
\end{align}
{ In order to simplify the computation of 
the power spectrum we utilize the initial states
from the standard inflationary scenario and we plug the
LQC effective equations for the background.}

Finally, we would like to compare our analytical results with 
those of previous studies. In particular, Herrera \cite{Herrera:2010yg} 
studied the canonical warm inflationary 
model in LQC (for similar studies see \cite{Herrera:2013rra, Zhang:2013yr,Xiao:2011mv,Herrera:2014mca,Xiao:2014nda}) under specific conditions, namely
(i) the potential has the form 
$V(\phi) \propto \phi^{2}$ (chaotic potential)
and (ii) the dissipation rate $\Gamma$ is constant. 
Also, Herrera et al. \cite{Herrera:2013rra} extended 
the analytical solutions of Ref.\cite{Herrera:2010yg} by taking into 
account a general form of $\Gamma$. However, the aforementioned studies
are in the context of 
standard scalar field theory which means that 
the density and the corresponding pressure are given by 
$\rho_{\phi}={\dot \phi}^{2}/2+V(\phi)$ and $p_{\phi}={\dot \phi}^{2}/2-V(\phi)$.
In the current article we investigate, for a first time,  
the warm LQC-tachyon inflationary scenario [see Eq.(\ref{1.2}] 
for a large family of $\Gamma$ parametrizations 
($\Gamma \propto T^{m}/\phi^{m-1}$) and we provide the corresponding class 
of potentials [see Eq.(\ref{wr3})].

\section{ Temperature at the end of inflation:}
%\section{Temperature at the end of inflation}

In this section, following the methodology of \cite{Mielczarek:2010ag}
we attempt to derive the temperature at the end of warm inflationary scenario.
The entire cosmological history contains the following eras:
I) from $t_{k}$ (Hubble crossing time) till the end of 
slow-roll warm inflation which is denoted by $t_{end}$,
II) from $t_{end}$  (recombination era) 
till the recombination epoch which is indicated by $t_{rec}$ and 
III)  from $t_{rec}$ up to present time $t_{0}$, for which we 
have the matter and dark energy dominated eras. 
Therefore, it is easy to write 

\begin{align}\label{sx1}
e^{N_{tot}}=\frac{a_{0}}{a_{k}} =\frac{1}{a_{k}}= \frac{H_{k}}{k} = 
\frac{a_{0}}{a_{rec}}\frac{a_{rec}}{a_{end}}\frac{a_{end}}{a_{k}}=e^{N_{0}} e^{N_{rec}} e^{N}
\end{align}
and thus $N_{tot}=N_{0}+N_{rec}+N$. 
Notice that in the above expression 
$a_{0}\equiv 1$ is the scale factor at the present time, 
$k=a_{k}H_{k}$ is the Fourier mode, $N_{tot}$ is the total number of e-folds, 
$N_{rec} \equiv \frac{a_{rec}}{a_{end}}$, 
$N_{0} \equiv \frac{a_{0}}{a_{rec}}$ and $N$ is given by 
Eq.(\ref{cc1}). 

At the end of warm inflation, the cosmic expansion enters 
in the radiation dominated era in which the universe is 
full of relativistic particles. 
Under adiabatic circumstances the 
radiation entropy per comoving volume reads
\begin{align}\label{tt}
S = \frac{2 \pi^{2}}{45} g T^{3} a^{3}
\end{align}
from which we find 
\begin{align}
\frac{a_{rec}}{a_{end}} = \frac{T_{end}}{T_{rec}} \left(\frac{g_{end}}{g_{rec}}\right)^{\frac{1}{3}}
\end{align}
or
\begin{align}
\label{tt1}
e^{{N}_{rec}} = 
\frac{T_{end}}{T_{rec}} \left(\frac{g_{end}}{2}\right)^{\frac{1}{3}}.
\end{align}
In the last step we utilized the entropy conservation law of the 
adiabatic radiation phase, which means that
$g_{end}T^{3}_{end}a^{3}_{end}=g_{rec}T^{3}_{rec}a^{3}_{rec}$ 
and we have set $g_{rec}= g_{\gamma} =2$. 
On the other hand, the temperature at the recombination epoch satisfies the 
well known formula
\begin{align}
 T_{rec} = \left(1+ z_{rec}\right) T_{CMB}.
\end{align}
Inserting the above in Eq.(\ref{tt1}) we find 
\begin{align}\label{fg1}
e^{N_{0}}e^{ N_{rec}} =  
\frac{T_{end}}{T_{CMB}} \left(\frac{g_{end}}{2}\right)^{\frac{1}{3}},
\end{align}
where
\begin{align}
e^{N_{0}}=\frac{a_{0}}{a_{rec}} = \left(1+ z_{rec}\right).
\end{align}

Lastly, using Eqs. (\ref{wr1}, \ref{mm1}, \ref{wr7}), the Hubble 
parameter at the Hubble crossing time is written as a function 
of $N$, namely
\begin{align}\label{fg4}
H_{k}(N) = A f \left(I(N)\right)^{f-1}.
\end{align}
Substituting Eq. (\ref{fg1}) in Eq. (\ref{sx1}) and 
utilizing Eq. (\ref{fg4}) we obtain the 
temperature at the end of warm inflation 
\begin{align}\label{tend}
T_{end} = T_{CMB} \frac{A f \left(I(N)\right)^{f-1}}{k} 
\left(\frac{2}{g_{end}}\right)^{\frac{1}{3}} e^{-N}.
\end{align}
Obviously, in order to compute $T_{end}$ 
we need to know the parameters $A$, $f$ and $N$ or equivalently 
the pair $(n_{s},r)$. Of course, we expect that the 
temperature at the end of warm inflation to satisfy the 
following inequality $T_{BBN} < T_{end} <T_{I}$, where 
$T_{BBN}\sim 10^{-2}$Gev is the temperature at the Big Bang
nucleosynthesis (BBN) and $T_{I}$
is the estimated temperature scale of 
inflation, an upper bound 
of which is provided by Planck team \cite{Ade:2015lrj}, 
$\sim 10^{16}$Gev. 
%{\bf QUESTION:what is the connection between this scale with that of $H_I \sim 10^{17}$Gev
%provided in section II. We need to understand that.
%The answer is that the energy density scale of Super-inflation is bigger than the energy density scale of normal inflation.}

%which in inflaton-type models is related to the – approximately constant– scalar potential during inflation

\section{ Observational constraints:}
%\section{Observational constraints}
%Let us finalize the paper by constraining the model at hand utilizing Plack 2015 results. Through this process, we strongly impose three constraints on the parameters of the model, namely $A$ and $f$; first, utilizing the results from Planck 2015, we can search for the adequate values of parameters of the model whereby the correponding value of spectral index $n_{s}$ and tenso-to-scalar ratio $r$ locate inside $r-n_{s}$ plane. The second condition arises as the temperature of inflation during slow-roll phase should be more than Hubble scale ($T > H$). Therefore, we attempt to find for which achieved values of $(A, f)$ in first stage, the second condition is also satisfied. Finally, following the formula has been derived for the temperature at the end of inflation and the lower and upper bound coming from observations, we can figure out for which values of $(A, f)$ satisfying first and second condition, the temperature at the end of inflation situates inside the range imposed by observations.
In order to check the consistency of the above slow-roll predictions 
with observation, we compare our results against those of 
\textit{Planck}  \cite{Ade:2015lrj}. 
Notice, that the pair $(n_{s},r)$ is given by Eqs.(\ref{sz1},\ref{sz11}).
Also, regarding the number of e-folds we use $N=50$ and $N=60$ respectively.

In particular we find: 
%Bellow we present our predictions 
%in the following cases:
\begin{itemize}

\item For the $\Gamma_{-1}$ ($m=-1$) parametrization: 
In fig.(\ref{m-1}) we present the $(n_{s},r)$ contours
together with our prediction for various 
values of $f$, $A$ and $C_{\phi}$.
Specifically, 
%Using perturbation parameter, we can compare this case 
%with observational data (see fig.(\ref{m-1})). 
%With the special choices of $f, A$ and $C_{\phi}$, we 
%have plotted  fig.(\ref{m-1}). 
in this graph the solid, dot-dashed and dashed lines 
correspond to $(A,f,C_{\phi})=(3.1, 0.7,10000)$, 
$(A,f,C_{\phi})=(2.8, 0.63,5000)$ and 
 $(A,f,C_{\phi})=(4, 0.63,15000)$ respectively.
We observe that in the case of large $C_{\phi}$ 
the scalar-to-tensor ratio $r$ (\ref{vv2}) 
becomes small with respect to the reference Planck result, while 
the spectral index lies in the interval $0.955<n_s<0.975$
%The opposite holds for the scalar index $n_{s}$ (\ref{sz1}) 
%({\bf NEW; IS IT CORRECT? PLEASE CONFIRM}).
Therefore, we argue that it is always possible to find the 
appropriate value of 
$C_{\phi}$ in order to get predictions which are in agreement
with those of 
Planck. 
For example, using $A=3.3, f = 0.7, C_{\phi}=15000$ we obtain 
$(n_{s},r)=(0.965691, 0.0141694)$.
\begin{figure}[h]
 \includegraphics[scale=0.6]{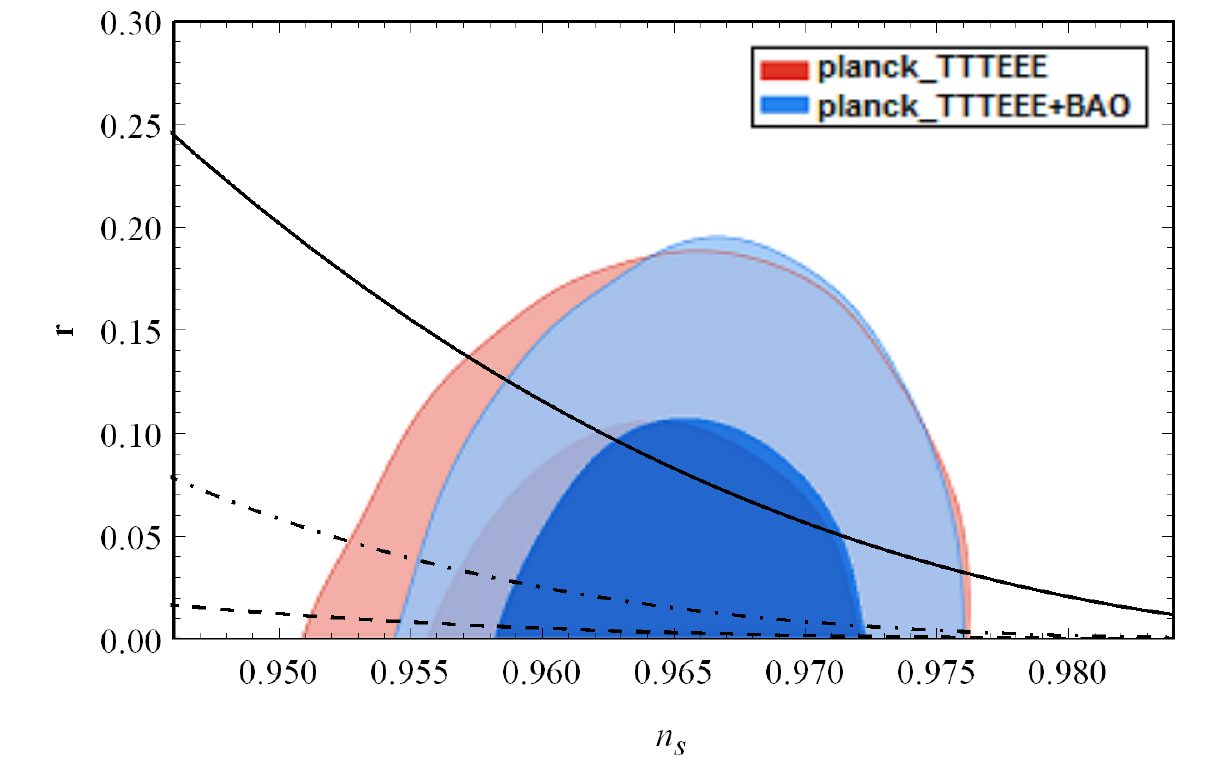} \ \ \ \ \ \ \ \ \ \  \includegraphics[scale=0.6]{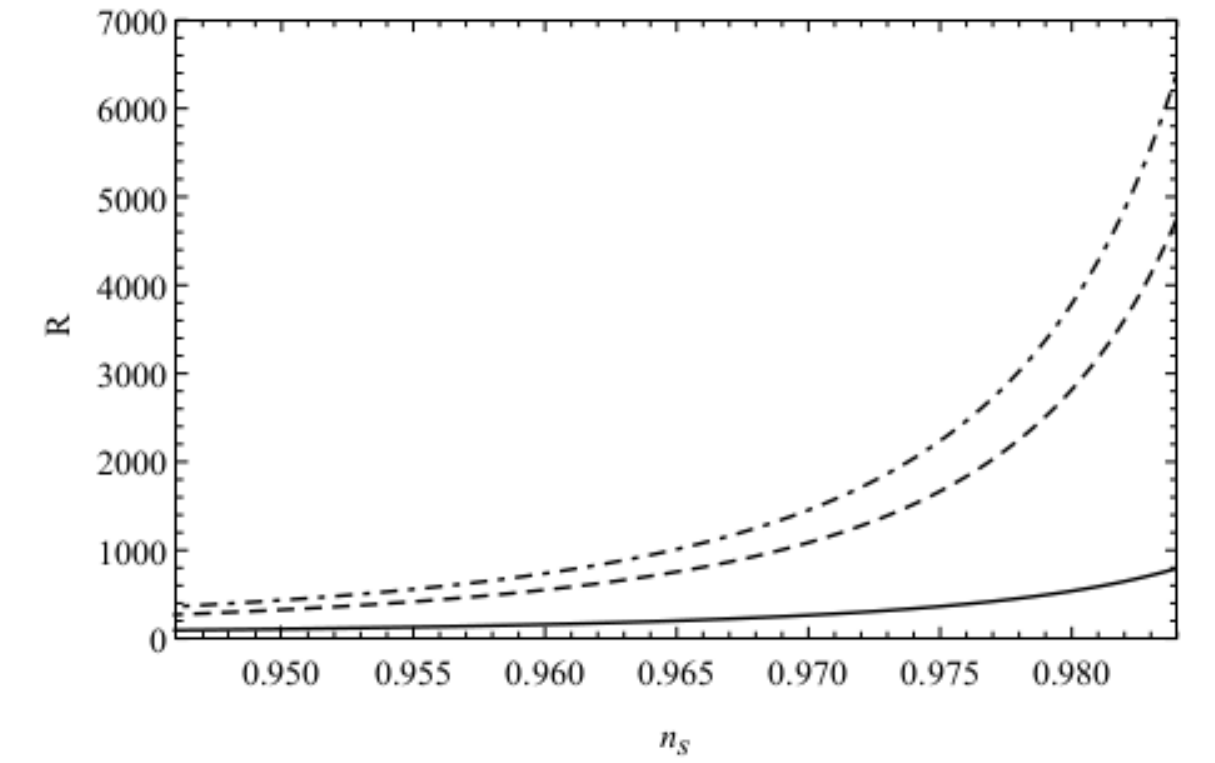} \\ % \includegraphics[scale=0.6]{rhorhoc123m-1} \ \ \ \ \ \ \ \ \ \   \includegraphics[scale=.6]{TH123m-1}
\caption{$1\sigma$ and $2\sigma$ contours borrowed from Planck  \cite{Ade:2015lrj}.
Our results are given in the case of $\Gamma_{-1}$ ($m=-1$) 
parameterization. Notice, that solid, dotdashed and dashed curves denote $(A= 3.1, f = 0.7, C_{\phi} = 10000)$, $(A = 2.8, f = 0.63, C_{\phi}  =5000)$ and $(A = 4, f = 0.63, C_{\phi} = 15000)$ respectively.} 
\label{m-1}
\end{figure}

%For example, using $C_{-1}=XXXX$ we obtain 
%$(n_{s},r)=(XXX,XXX)$. ({\bf INTRODUCE THE APPROPRIATE VALUES})

\item For the $\Gamma_{0}$ ($m=0$) parametrization: 
here our results can be found in figure (\ref{m0}) 
for the following special cases:   
$(A, f,C_{\phi})=(4.6,0.59,10000)$ [see solid curve], 
$(A, f,C_{\phi})=(3.1,0.59,5000)$ [see dashed curve], 
and $(A, f,C_{\phi})=(4.3,0.59,15000)$ [see dot-dashed curve]. 
As in the previous case, also here we need to fine-tune 
$C_{\phi}$ in order to be consistent with observations. 
As an example, using $A=4.6, f = 0.59, C_{\phi}=15000$ we obtain 
$(n_{s},r)=(0.960205, 0.0191892)$.
\begin{figure}[h]
 \includegraphics[scale=0.6]{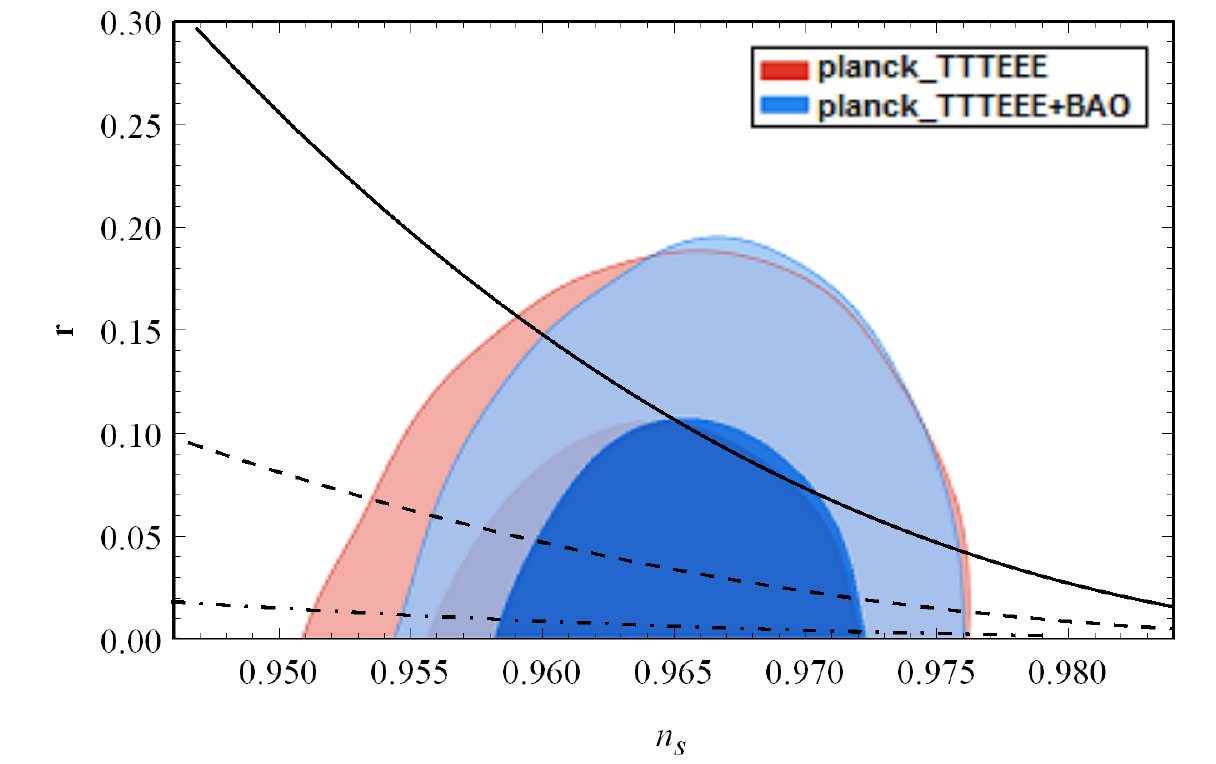} \ \ \ \ \ \ \ \ \ \  \includegraphics[scale=0.6]{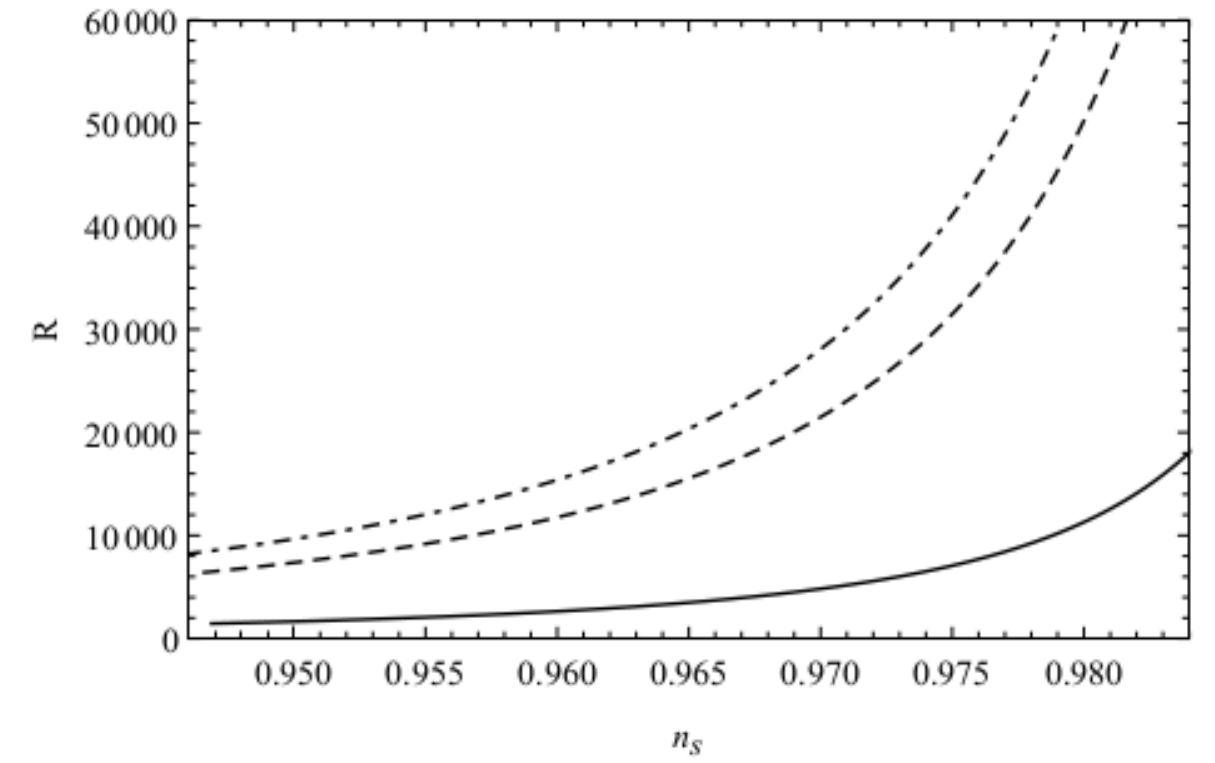} \\  %\includegraphics[scale=0.6] {rhorhocm0} \ \ \ \ \ \ \ \ \ \   \includegraphics[scale=.6]{TH123m0}
\caption{The same contours as in figure 1. Here our 
results corresponds to $\Gamma_{0}$ ($m=0$) parametrization.
The dashed, solid and dot-dashed curves denote $(A= 3.1, f = 0.59, C_{\phi} = 5000)$, $(A = 4.6, f = 0.59, C_{\phi}  =10000)$ and $(A = 4.3, f = 0.59, C_{\phi} = 15000)$ respectively.}
\label{m0} 
\end{figure}
\item For the $\Gamma_{1}$ ($m=1$) parametrization:
We argue bellow that this model 
alleviates the fine-tuning issue of the above parametrizations.
We remind the reader warm inflation satisfies the following 
condition $T>H$ (or $\frac{T}{H}>1$). 
%It is clear that $m=1$ case is the most suitable one to be 
%consistent with observation without fine tuned value of 
%parameter $C_{\phi}$. For this reason we proceed our extended 
%analysis considering only $m=1$ case. 
%On of the important condition 
%which must be satisfied is $\frac{T}{H}>1$.
For $m=1$ the ratio between the $T$ and $H$ 
%({\bf NEW PROVIDE THE REFERENCE NUMBER OF EQUATIONS FOR H AND T. This 
%will help the reader to extract eq.63})
is given by (\ref{tem}),(\ref{sd1}),(\ref{cc1}): 
\begin{equation}\label{eq:t-h}
	\frac{T}{H} = \left(\frac{3}{2C_{\gamma}}\right)^{\frac{1}{4}}
\left(\frac{1-f}{f^3A^3}\right)^{\frac{1}{4}}(I(N))^{\frac{3f-2}{4}}
\end{equation} 
In Fig.(\ref{fig:t-h}) we show $\frac{T}{H}$ in $A-f$ plane. 
The black solid line corresponds to the boundary limit 
$\frac{T}{H}=1$, while the top left part of 
the plane is consistent with the restriction $\frac{T}{H}>1$. 
%In fact only these 
%$A-f$ pairs are consistent with warm inflation condition.
Moreover, in Fig.(\ref{fig:cons}) we plot the $A-f$ allowed area 
in which our $(n_{s},r)$ results are in agreement with 
those of Planck within $1\sigma$ errors.
Notice, that the transparent background 
(foreground opaque) corresponds to $N=60$ ($N=50$).
From this figure we observe  
that for different values of the dissipation coefficient 
$C_{\phi}$ we always find a narrow strip in the $A-f$ parameter 
space which is in agreement with Planck's priors of 
$n_{s}$ and $r$. 
%are not very different in this case. 
Indeed, in Fig.(\ref{fig:planck}) we plot 
the Planck confidence contours in the plane of $(n_{s},r)$. 
On top of that we show the big solid point for the individual 
pair of $(n_{s},r)$ in the case of $N=60$. Also, 
we provide the corresponding small solid point for $N=50$.
%The comparison indicates that we can always find a
%$A-f$ parameter space for which the warm inflationary LQC scenario 
%within the context of $\Gamma_{1}$ parametrization 
%for some values of our parameters$(f,A)$} is 
%consistent with Planck data.
%that the predicted 
%$(n_{s},r)$ Planck
%data favor 
%the warm inflationary LQC scenario 
%within the context of $\Gamma_{1}$ parametrization {\bf for some values of our parameters$(f,A)$}. 
\begin{figure}
	\centering
	\includegraphics[width=0.48\textwidth]{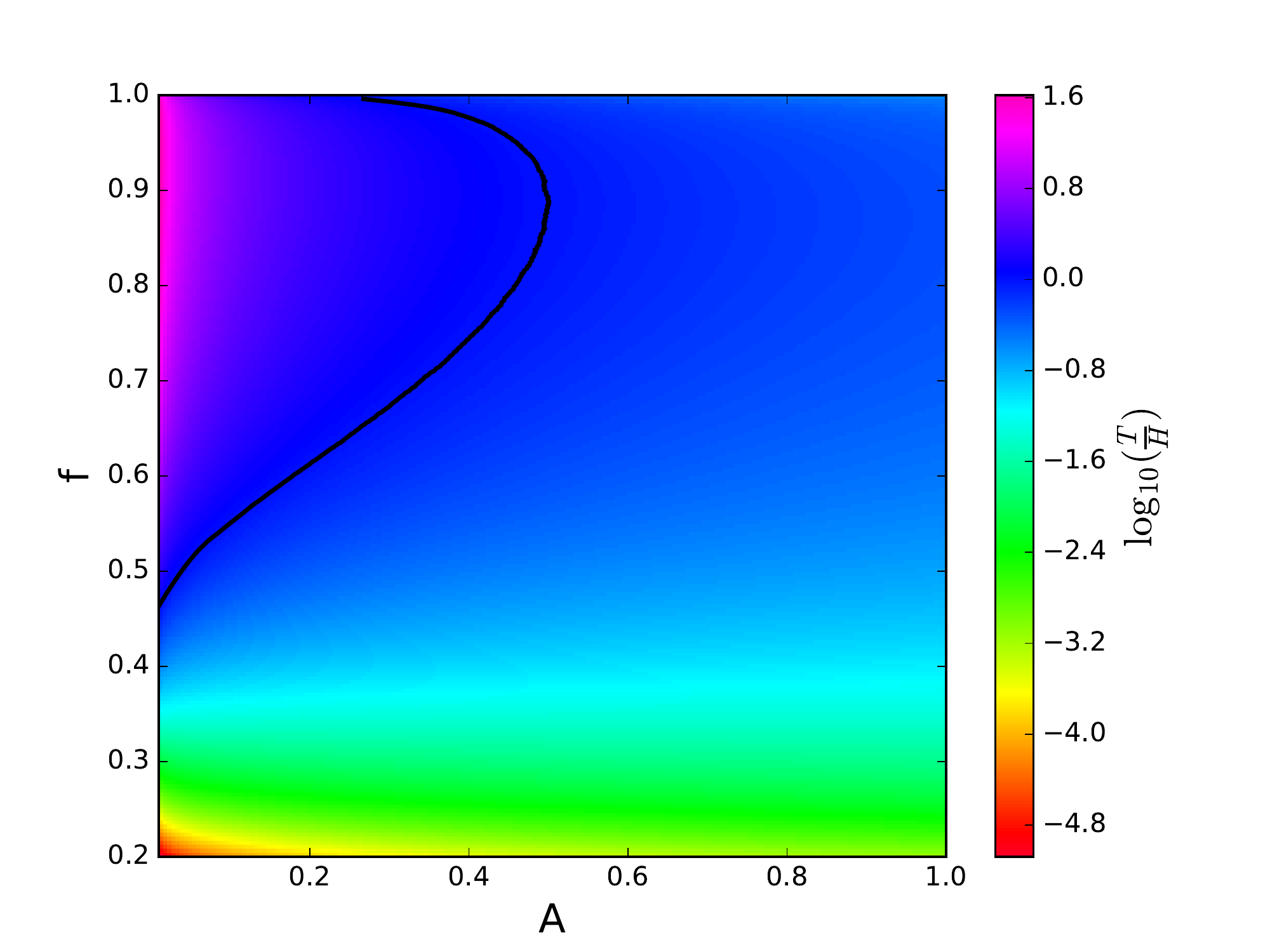}
	\caption{The ratio $\frac{T}{H}$ in the A-f plane. 
The black solid line indicates the boundary $\frac{T}{H}=1$ and the 
left top part of plane corresponds to $\frac{T}{H}>1$. }
	\label{fig:t-h}
\end{figure}

\begin{figure}
	\centering
	\includegraphics[width=0.48\textwidth]{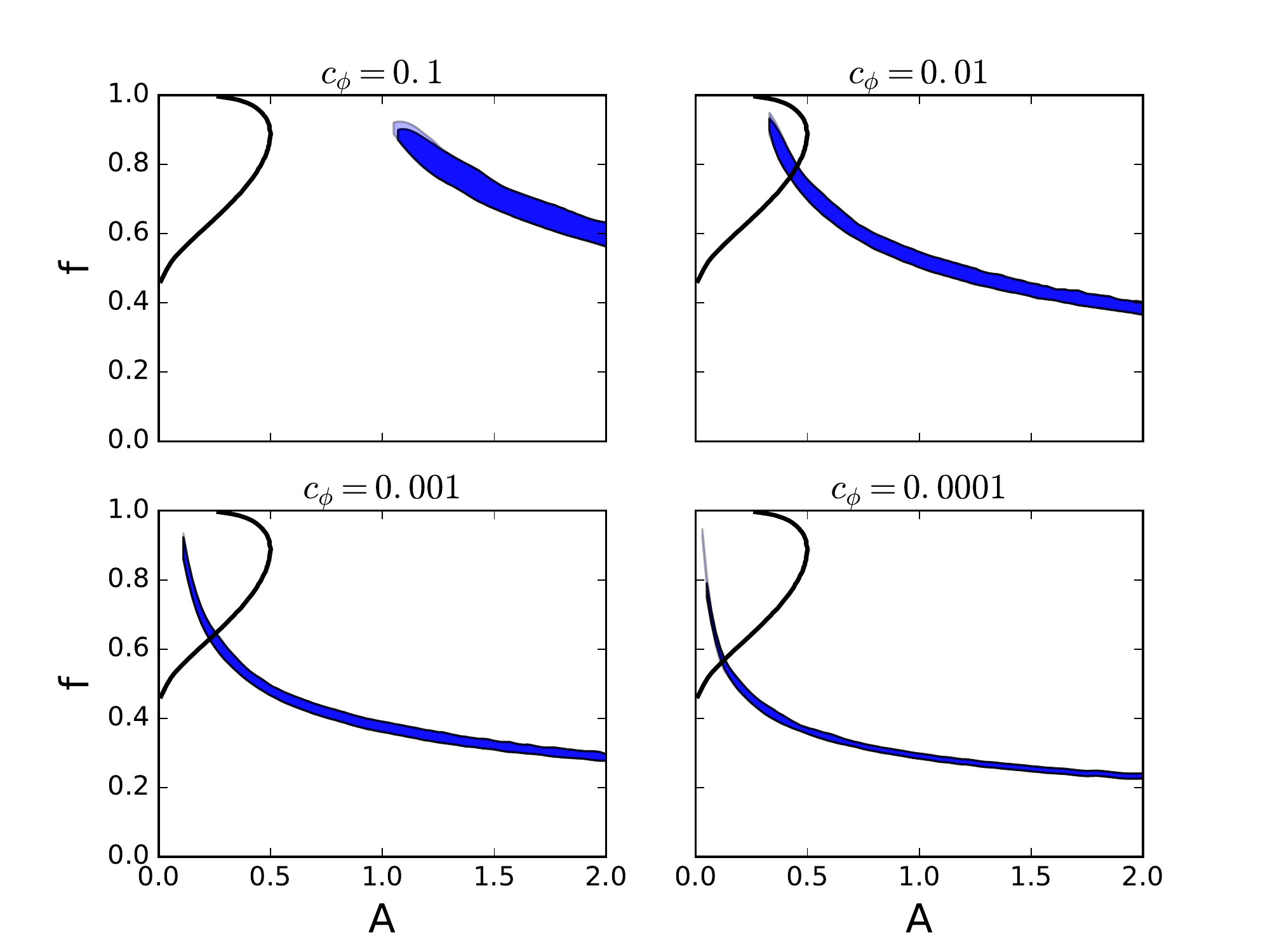}
	\caption{A-f pairs which are consistent within $1\sigma$ region of \em{Planck} results. In each panel the $C_{\phi}$ is given in top panel and the black solid line is same as Fig.(\ref{fig:t-h}).}
	\label{fig:cons}
\end{figure}

\begin{figure}
	\centering
	\includegraphics[width=0.48\textwidth]{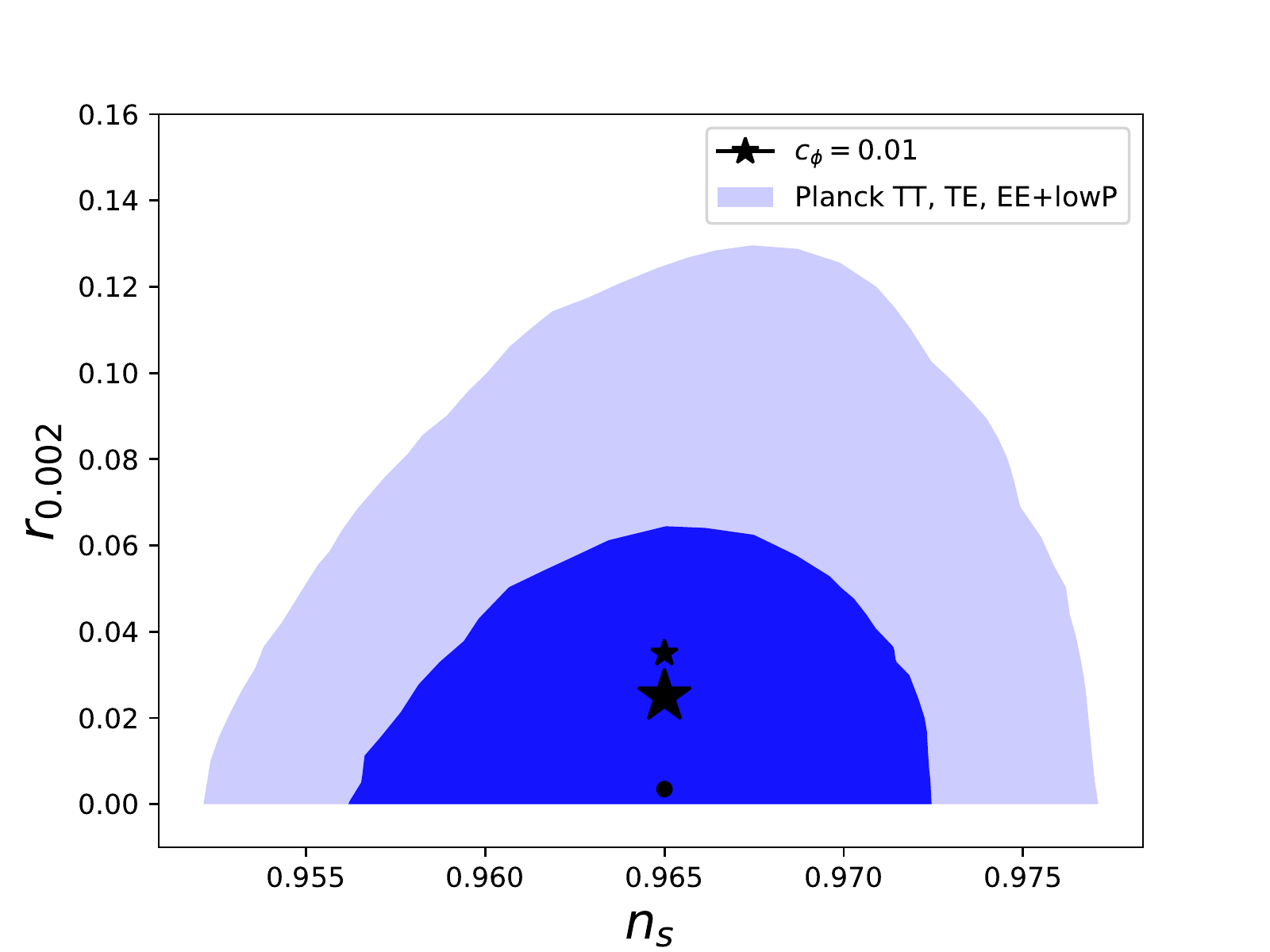}
	\caption{$1\sigma$ and $2\sigma$ confidence regions of \em{Planck} result. The big (small) star indicates N=60(N=50) e-folding of warm LQC-tachyon inflation model and the disk indicates N=60 for non-LQC warm Scenario. Here we use $C_{\phi}=0.01$, $f=0.9$ and $A=0.35$ for case $m=1$.}
	\label{fig:planck}
\end{figure}
\end{itemize}

It is worth noting that concerning the observational signatures of LQC 
in the CMB data, an intense debate is taking place in the 
literature about the implementation of LQC to CMB data. 
Recently, Ashtekar and Gupt (see Ref.\cite{Ashtekar:2016wpi})
using various correlation functions for scalar perturbations 
found that LQC is favored by Planck, while standard
(cold) inflation can not accommodate the data at large angular scales 
($l \le 30$).
The heated discussion is going and the aim of our study 
is to contribute to this debate.
Within this framework in figure 5 we plot 
the predicted $(n_{s},r)$ in the case non-LQC warm inflation
(see stars in figure 5).
We then compare the latter $(n_{s},r)$ predictions with those
of the  warm LQC-tachyon inflation (solid points). 
In principle, this can help us to understand better the theoretical 
expectations of the  warm LQC-tachyon inflationary model, as 
well as to identify the differences from 
the non-LQC warm tachyon inflation. From figure 5 we observe that 
both inflationary tachyon models provide the same 
spectral index $n_{s}$. Concerning the tensor-to-scalar fluctuation ratio $r$
the situation is different.
Although the predictions are in agreement 
with Planck observations (within 1$\sigma$), we find that the 
non-LQC warm tachyon 
inflation provides 
a tensor-to-scalar fluctuation ratio which is smaller than that 
of  warm LQC-tachyon inflation. Therefore, in the light of the next generation 
of $B-$mode data one may use 
this difference in order to test the performance 
of LQC in the CMB data. 

%\textbf{ We can distinguish the spectrum of B-modes of our super-inflation model and standard inflation using future observational data \cite{Ribassin:2011km}. In Fig.\ref{fig:planck} we can see the standard model of warm inflation has lower value than the warm inflation in the context of LQC model which is result of reducing of $\dot{\phi}^2$ in standard model of inflation (see Eqs.(\ref{sm1},\ref{vv2}))}  

%\begin{figure}[H]
% \includegraphics[scale=0.6]{r123m0} \ \ \ \ \ \ \ \ \ \  \includegraphics[scale=0.6]{q123m0} \\  %\includegraphics[scale=0.6] %{rhorhocm0} \ \ \ \ \ \ \ \ \ \   \includegraphics[scale=.6]{TH123m0}
%\caption{The results have been plotted for $m=0$ in which dashed, solid and dotdashed curves denote $(A= 3.1, f = 0.59, C_{0} = %5000)$, $(A = 4.6, f = 0.59, C_{0}  =10000)$ and $(A = 4.3, f = 0.59, C_{0} = 15000)$ respectively.}
%\label{m0} 
%\end{figure}

\section{Conclusions:}
%\section{Conclusions}
In this work, 
we studied the observational signatures of quantum cosmology 
in the Cosmic Microwave Background data given by Planck2015.
We utilized the paradigm of warm inflation with a tachyon scalar field 
to the loop quantum cosmology. Within this framework, we first 
derived the main cosmological quantities as a function of 
the tachyon field. 
Second, we provided the slow-roll parameters and the 
power spectrum of scalar and tensor fluctuations respectively.
Finally, we checked the performance of various warm inflationary models 
against the data provided by Planck2015 data and we find a class 
of patterns which are consistent with the observations. 

%In this  work we investigated the tachyon inflation on the brane in the context of 
%a spatially flat Friedmann-Robertson-Walker  universe. We adopted a specific form of scale factor
%from Barrow \cite{Barrow:1996bd} solutions, namely logamediate scale  factor.
%Within this context, we estimated analytically the  
%slow-roll parameters potential of the model and compare  predictions  
%with those of other famous inflationary models in the literature.
%Confronting the model against
%the latest observational data, we found that the tachyon 
%inflationary model on the brane  
%is consistent with the results presented in \emph{Planck 2015} within
%$1\sigma$ uncertainties.
{\bf Acknowledgment:} 
%\section{acknowledgement}
%\begin{thebibliography}{999}
%\bibitem{general dissipative coefficient}Y. Zhang, JCAP {\bf0903}, 023 (2009) \href{https://arxiv.org/abs/0903.0685}{hep-ph/0903.0685}
%\bibitem{temperature at the end of inflation} J. Mielczarek, Phys. Rev. D {\bf 83}, 023502 (2011), \href{https://arxiv.org/abs/%1009.2359}{astro-ph/1009.2359}
%\end{thebibliography}
SB acknowledges support by the Research Center 
for Astronomy of the Academy of Athens in the
context of the program ''Testing general relativity on cosmological scales''
(ref. number 200/872).

 \bibliographystyle{apsrev4-1}
  \bibliography{ref}
\end{document}